\documentclass[sigconf]{acmart}

\usepackage{comment}
\usepackage{listings}

\usepackage{caption}
\usepackage{subcaption}

\newcommand{\insitu}{\textit{in situ}}
\newcommand{\intransit}{\textit{in transit}}

\definecolor{codegreen}{rgb}{0,0.6,0}
\definecolor{codegray}{rgb}{0.5,0.5,0.5}
\definecolor{codepurple}{rgb}{0.58,0,0.82}
\definecolor{backcolour}{rgb}{0.95,0.95,0.92}
\definecolor{codecyan}{rgb}{0.0,0.2,1.0}

\lstdefinestyle{mystyle}{
    commentstyle=\textcolor{codegreen},
    keywordstyle=\color{codecyan},
    numberstyle=\tiny\color{codegray},
    stringstyle=\color{codepurple},
    basicstyle=\ttfamily\footnotesize,
    breakatwhitespace=false,    
    breaklines=true,    
    captionpos=b,    
    keepspaces=true,    
    numbers=left,    
    numbersep=2pt,  
    firstnumber=auto,
    numberblanklines=false,
    showspaces=false,
    showstringspaces=false,
    showtabs=false,
    tabsize=2
}
\AtBeginDocument{%
  }

\setcopyright{acmcopyright}
\copyrightyear{2023}
\acmYear{2023}
\acmDOI{XXXXXXX.XXXXXXX}

\acmConference[Conference acronym 'XX]{Make sure to enter the correct
  conference title from your rights confirmation emai}{June 03--05,
  2018}{Woodstock, NY}
\acmPrice{15.00}
\acmISBN{978-1-4503-XXXX-X/18/06}

\begin{document}

%%
%% The "title" command has an optional parameter,
%% allowing the author to define a "short title" to be used in page headers.
\title{Towards a Scalable In Situ Fast Fourier Transform}

%%
%% The "author" command and its associated commands are used to define
%% the authors and their affiliations.
%% Of note is the shared affiliation of the first two authors, and the
%% "authornote" and "authornotemark" commands
%% used to denote shared contribution to the research.

\author{Sudhanshu Kulkarni}
\affiliation{
    \institution{San Francisco State University}
    \city{San Francisco, CA}
    \country{USA}
    }
\email{skulkarni@sfsu.edu}

\author{Burlen Loring}
\affiliation{
    \institution{Lawrence Berkeley National Laboratory}
    \city{Berkeley, CA}
    \country{USA}
    }
\email{bloring@lbl.gov}

\author{E. Wes Bethel}
\orcid{0000-0003-0790-7716}
\affiliation{
    \institution{San Francisco State University}
    \city{San Francisco, CA}
    \country{USA}
    }
    \affiliation{
    \institution{Lawrence Berkeley National Laboratory}
    \city{Berkeley, CA}
    \country{USA}
    }
\email{ewbethel@sfsu.edu}

%%
%% By default, the full list of authors will be used in the page
%% headers. Often, this list is too long, and will overlap
%% other information printed in the page headers. This command allows
%% the author to define a more concise list
%% of authors' names for this purpose.
\renewcommand{\shortauthors}{Kulkarni et al.}

%%
%% The abstract is a short summary of the work to be presented in the
%% article.

\begin{abstract}
    The Fast Fourier Transform (FFT) is a numerical operation that transforms a function into a form comprised of its constituent frequencies and is an integral part of scientific computation and data analysis.
    The objective of our work is to enable use of the FFT as part of a scientific \insitu{} processing chain to facilitate the analysis of data in the spectral regime.
    We describe the implementation of an FFT endpoint for the transformation of multi-dimensional data within the SENSEI infrastructure.
    Our results show its use on a sample problem in the context of a multi-stage \insitu{} processing workflow.

\end{abstract}

%%
%% The code below is generated by the tool at http://dl.acm.org/ccs.cfm.
%% Please copy and paste the code instead of the example below.
%%



%%
%% Keywords. The author(s) should pick words that accurately describe
%% the work being presented. Separate the keywords with commas.

% \received{20 February 2007}
% \received[revised]{12 March 2009}
% \received[accepted]{5 June 2009}

%%
%% This command processes the author and affiliation and title
%% information and builds the first part of the formatted document.
\maketitle

\begin{figure*}
\centering
\includegraphics[width=\textwidth]{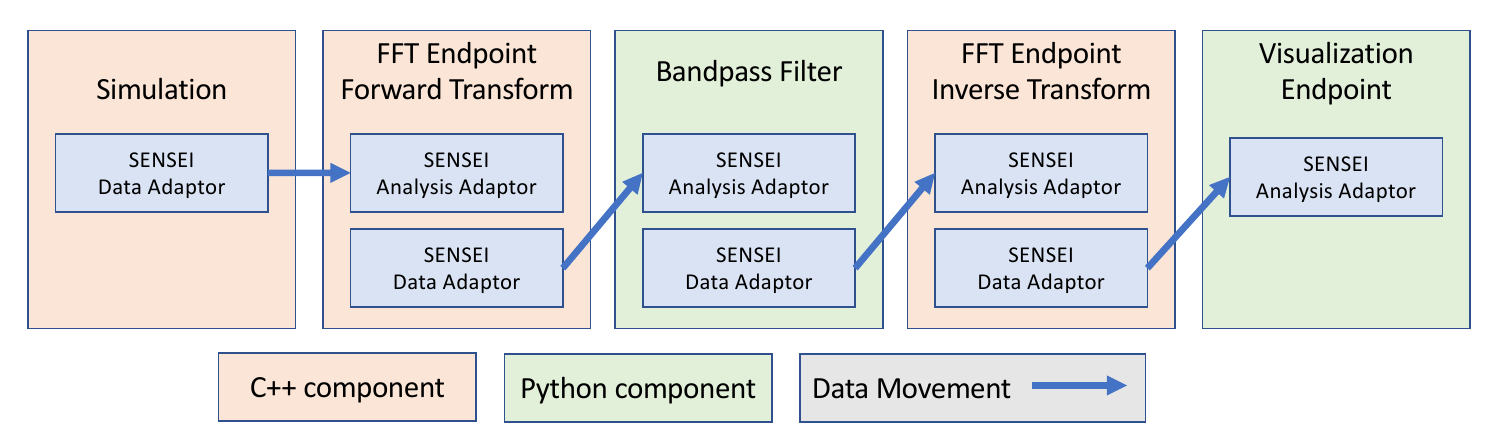}
\caption{High-level component diagram of the FFT Endpoint.}
\label{fig:components_diagram}
\end{figure*}
\section{Introduction}
%This is the introduction.

%Problem definition: what problem are we trying to solve? Why is this a problem worthy of solving?

The Fourier Transform (FT) is a numerical operation that transforms a function or field into a form that describes the weights of frequencies which when summed together describes the original field. 
The Fast Fourier Transform (FFT), which is a specialized implementation of the FT, is considered to be one of the top 10 algorithms of the 20\textsuperscript{th} century and is widely used by applications in science and engineering~\cite{heFFTe:2020}.
While there exist high-quality FFT implementations like FFTW~\cite{Frigo:FFTW:2005} and heFFTe~\cite{heFFTe:2020}, these are accessible via a library API.
The focus of our work here is to leverage this previous work and bring these capabilities into the \insitu{} environment thereby enabling FFT-based methods for use on HPC platforms without having to first make a round trip to persistent storage.

%Couple of sentences of previous work in scalable FFT, and any previous work about in situ FFT (there probably is none).

Our approach is to leverage the FFTW implementation~\cite{Frigo:FFTW:2005} as the underlying scalable FFT implementation in our system.
We create the infrastructure around FFTW so that this new FFT capability is accessible in the SENSEI infrastructure~\cite{Ayachit:SENSEI-interface:2016}, particularly as part of multi-stage \insitu{} processing chains on HPC platforms. 

The design objectives for the work in this document focus on supporting forward and inverse transforms and their use as part of a heterogeneous processing chain in either \insitu{} or \intransit{} processing scenarios. 
%\fix{Add back in the 1D/2D/3D data if we have some 1D examples to show, otherwise comment out this sentence.}
%, and support for 1D/2D/3D structured scientific data.
Longer term objectives include scalable \insitu{} workloads where producer and consumer are both running at scale and at concurrency levels commensurate with memory footprint requirements for larger problem sizes.

Our implementation is an FFT endpoint that is part of the SENSEI \insitu{} infrastructure. As such, the FFT endpoint be combined with other SENSEI tools in a daisy-chain fashion to create a multi-stage workflow. The FFT endpoint, like other SENSEI endpoints, is configured via an XML file so that its parameters may be manipulated at runtime through changes to the configuration file.
Our results demonstrate its use as part of a heterogenous \insitu{} processing chain that consists of a data producer, a forward FFT, operations on spectral-domain data, an inverse FFT, and visualization of results. 

The primary contribution of this work is the presentation of a new SENSEI \insitu{} FFT endpoint based upon the FFTW~\cite{Frigo:FFTW:2005} implementation.
This new FFT endpoint is suitable for use in a scientific \insitu{} processing chain consisting of heterogeneous processing stages. 
Such a capability is missing from the \insitu{} technology landscape and is broadly applicable to many different scientific domains and science applications.

\section{Design and Implementation}
\label{sec:design}

% ===================
\subsection{Design Objectives}
\label{sec:design:objectives}

% Enumerate the design objectives. We can think about near-term and longer-term objectives.

% Near term: data dimensions. Want to be able to do forward, inverse FFT on 1D, 2D, and 3D datasets.

% Near term: want to do forward, inverse FFTs as part of an in situ processing chain. This is the primary use case.

% Longer-term: MxN in situ use case. 

The problem we are solving is to enable the use of the FFT as part of a heterogeneous \insitu{} workflow consisting of multiple processing stages each of which may be implemented in different ways.
A prototypical use case is where a data generator, like a simulation, produces data that is then processed by downstream components that may be configured for use in either an \insitu{} or \intransit{} configuration.
Fig.~\ref{fig:components_diagram} illustrates this idea using components having varying implementations. 

%This work focuses on enabling the use of the FFT as part of a processing chain specifically for either \insitu{} or \intransit{} deployment.
%
%In that context, we want  ... \fix{Wes to finish this thought}

The prime motivation in designing this system is to provide a single FFT endpoint that can be connected directly to a simulation. As a near-term objective, we want to be able to do a Forward and Inverse FFT on a multi-dimensional dataset as a part of an \insitu{} processing chain. This design will evolve further to support advanced load distribution techniques for highly optimized parallel and scalable FFT-analysis implementations.

% ===================
\subsection{The SENSEI FFT Endpoint}
\label{sec:design:fft_endpoint}

In order to enable the use of the underlying FFTW implementation, one of the primary tasks of the SENSEI FFT endpoint is to manage data marshaling between the SENSEI bridge data model and the FFTW data model.
Such marshaling will transform data from the SENSEI bridge into FFTW for either forward or inverse transformations, and then from FFTW back into the SENSEI bridge for use by other SENSEI-enabled components that may be part of an \insitu{} processing chain.
The FFTW data model consists of either real or complex-valued structured meshes in 1D, 2D, or 3D.

SENSEI's bridge data model~\cite{Ayachit:SENSEI-interface:2016}  is based upon the VTK data model due to its broad applicability to many different science data types and its use in applications like VisIt and ParaView. 
Codes that produce data for use by other \insitu{} components will make use of the SENSEI Data Adaptor to map from their source data model into the SENSEI bridge data model.
Codes that consume data produced in an \insitu{}
context will use SENSEI's Analysis Adaptor to map from the SENSEI bridge into the local data model.

%\fix{Wes to write a few sentences here about the FFTW data model. My guess is that it is a 1D/2D/3D structured mesh. }

\subsubsection{SENSEI FFT Endpoint Configuration} \hfill  
  
To configure the FFT Analysis backend, we use the infrastructure provided by SENSEI's Configurable Analysis Adaptor. 
An XML file, like presented in Listing ~\ref{listing:fft-xml}, holds all the necessary parameters used for configuration. 
The \texttt{mesh} and \texttt{array} attributes are used to identify data objects in the Data Adaptor which contain the simulation data. The \texttt{direction} variable is used to toggle between forward and inverse FFT. 
To extend this pipeline further, the FFT analysis backend has provisions to send the data to a Python endpoint by employing the Python Analysis Adaptor, which will be configured via \texttt{python\_xml}.

\begin{lstlisting}[caption={A sample FFT configuration XML to invoke Forward FFT on the passed array and mesh.},label={listing:fft-xml}, name=fft-xml, float=h, style=mystyle, language=xml]
<sensei>
  <analysis type="fft" mesh="mesh" array="dataArray" direction="FFTW_FORWARD"  python_xml="python_spectral_config.xml"/>
</sensei>
\end{lstlisting}

% ===================
\subsubsection{SENSEI FFT Data Marshaling and Execution} \hfill
\label{sec:FFT:marshal_execute}

% Describe the objectives and code for going from the SENSEI bridge to FFTW data models. This is the SENSEI Analysis Adaptor.

% The idea is to transform from the SENSEI bridge so that FFTW can do a forward or inverse transform.

% What type of data and metadata are required and used as part of this process?

% What does the code look like for 1D, 2D, 3D data?

% \fix{Start out by describing the task: we are transforming data from the SENSEI bridge model into a form needed by FFTW, we are executing FFTW, then we are transforming FFTW output results into the SENSEI brdige for others to consume.}

% \fix{Explain how you access the simulation data and set things up for FFTW.}

% \fix{Show a few lines of your C++ code where you do these conversions and explain what they do.}

% \fix{Explain that you then invoke the FFTW method. This may take a few sentences because you have to create the plan and then execute the plan. Also, include a few code snippets to show off these results.}

% \fix{Then explain that next you have to transform from FFTW output into the SENSEI bridge model for others to consume.}

% \fix{Show some lines of code where you do this.}

The idea here is to transform the data model from the SENSEI bridge so that FFTW can perform its computations. We use three components for this transformation; the dimensions, the direction, and the data.
% \fix{In this context, what is "direction"? Do you mean specifying whether a forward or inverse FFT?}
The direction, which specifies whether to perform forward or inverse FFT, is provided by the XML file, while the rest of the components are extracted from the Data Adaptor.

The simulation must pass an instance of SENSEI Data Adaptor while triggering the \insitu{} processing. The process begins by fetching the requested data object from the simulation and adding data arrays from the simulation to this mesh object. This object is shipped as a \texttt{vtkMultiBlockDataSet}, which has to be cast into a \texttt{vtkDataObject} before we can get the actual array out of it. Details of this process can be found in Listing ~\ref{listing:get_data}. 

\begin{lstlisting}[caption={Extracting data from simulation},label={listing:get_data}, name=get_data, float=h, style=mystyle, language=c++]
// convert to svtkDataObject to svtkMultiBlockDataSet and process the blocks. (svtk is SENSEI's wrapper over vtk library)
svtkCompositeDataSetPtr mesh = SVTKUtils::AsCompositeData(Comm, dobj, true);
svtkSmartPointer<svtkCompositeDataIterator> iter;
iter.TakeReference(mesh->NewIterator());

// iterate over each block
for (iter->InitTraversal(); !iter->IsDoneWithTraversal(); iter->GoToNextItem())   
  svtkDataObject *curObj = iter->GetCurrentDataObject();

  // get array from the data object
  svtkFieldData* fd = curObj->GetAttributesAsFieldData(0))
  
  svtkDataArray* simulation_data = fd->GetArray(array_name.c_str());

\end{lstlisting}

% Using a \texttt{vtkSmartPointer}, we will iterate over all blocks and fetch the data array from it. Each block uses \texttt{vtkImageData} object to load the data, as it best suites our purpose as we are working with structured meshes. Dimensions are retrieved from the image data, as shown in Listing ~\ref{listing:get_dimensions}.

% \begin{lstlisting}[caption={Extracting dimensions of simulation data},label={listing:get_dimensions}, name=get_dimensions, float=h, style=mystyle, language=c++]
% // Cast the data object into its derived svtkImageData object and fetch dimensions from it
% svtkImageData* imageData = svtkImageData::SafeDownCast(curObj);

% int dims[3];
% imageData->GetDimensions(dims);    
% \end{lstlisting}

% \fix{Listing ~\ref{listing:get_dimensions} can be omitted if it feels too cluttered. It does not hold any significant code. Should we skip this listing - YES / NO. From ewb: my feeling is that listing ~\ref{listing:get_dimensions} doesn't add a lot to the story, so please delete it, along with the paragraph starting with "Using a vtkSmartPointer, ..." }

This \texttt{simulation\_data} can now be passed to the FFTW library for performing an \textit{in-place} transformation. For transforming the 2D dataset in our experiment, FFTW routines for parallel systems supporting the MPI message-passing interface are used. Listing ~\ref{listing:fftw} highlights the pseudo-code for this implementation.

% \fix{from ewb: here, what are you referring to "This data" at the start of this paragraph? YOu need to help the reviewers follow the story... towards the end of Listing~\ref{listing:get_data} you should have a variable that is clearly the VTK data object containing the simulation data, and then you need to carry that variable forward somehow into Listing~\ref{listing:fftw} so that it is easy to follow along what is happening. Also in Listing \ref{listing:fftw}, I'll suggest using slightly more descriptive variable names. A variable named "data" is too vague :) Maybe something like "fft\_input\_data" might be better: clearly identify inputs and outputs. }

\begin{lstlisting}[caption={Performing discrete FFT on data of size N0 x N1},label={listing:fftw}, name=fftw, float=h, style=mystyle, language=c++]
#include <fftw3-mpi.h>
fftw_plan plan;
fftw_complex *fftw_data;
ptrdiff_t alloc_local, local_n0, local_n0_start

fftw_mpi_init();

// Get local data size and allocate memory to 'fftw_data'
alloc_local = fftw_mpi_local_size_2d(N0, N1, MPI_COMM_WORLD, &local_n0, &local_n0_start);
fftw_data = fftw_alloc_complex(alloc_local);

// create a plan for in-place DFT
plan = fftw_mpi_plan_dft_2d(N0, N1, fftw_data, fftw_data, MPI_COMM_WORLD, direction, FFTW_ESTIMATE);  

// initialize fftw_data from the simulation_data
for(i = 0:local_n0) for(j = 0:N1)
  fftw_data[i*N1 + j] = simulation_data[local_0_start + i, j]

// compute transforms, in-place, as many times as desired
fftw_execute(plan);

fftw_destroy_plan(plan);
MPI_Finalize();

\end{lstlisting}
FFTW library follows a paradigm of allocate - plan - execute - destroy. It will request a few parameters while creating a plan for this; the dimensions, object of type \texttt{fftw\_complex} which holds the input data, the global MPI Communicator, the direction of transform intended (\textsc{fftw\_forward} or \textsc{fftw\_backward}), and a flag for notifying FFTW to use the computation model that is estimated to run in a reasonable time. The \texttt{simulation\_data} extracted from Data Adaptor will be copied into the \texttt{fftw\_data}, which is picked up by the plan. This plan, when executed computes the transform and stores it in the same data object.

The output will be packed in VTK data objects to be passed on the SENSEI Bridge for others to consume. This is achieved in a similar method as any simulation data might implement; a \texttt{vtkDoubleArray} to hold the data, passed via an \texttt{vtkImageData} to the Data Adaptor.

% ===================
% \subsubsection{SENSEI FFTW Data Adaptor} \hfill
% Describe the objectives and code for going from the FFTW data model to the SENSEI bridge model. This is the SENSEI Data Adaptor.

% The idea is to transform from FFTW to the SENSEI bridge so that the SENSEI FFTW adaptor may be used as part of an in situ processing chain.

% What type of data and metadata are required and used as part of this process?

% What does the code look like for 1D, 2D, 3D data?

% \fix{Sudanshu: this subsubsection is going away. I folded everything into Sec.~\ref{sec:FFT:marshal_execute} in an effort to streamline things and keep it as simple and straightforward as possible. Please transplant text from this section into that earlier section as needed to make your point.}

% The Data Adaptor is a key piece in this processing pipeline. The simulation must pass an instance of SENSEI DataAdaptor while triggering the in situ processing. It will hold a \lstinline{double} array of size $M\times N$ packed in mesh-based \lstinline{svtkMultiBlockDataObject}. Each block's data can be packed most efficiently in a \lstinline{svtkImageData} object. The extent of the object in it will determine the dimensions of the data in each block.

% Similarly, the FFT endpoint will configure the output data in the same fashion for further processing.

\subsection{Python-based Endpoints}
\label{sec:design:python_endpoint}

For our example shown in Fig.~\ref{fig:components_diagram} we have two Python-based components: one performs bandpass filtering to implement the noise reduction and one produces visualization.
These endpoints follow the design pattern described in Loring et al., 2018~\cite{loring2018pythonbasedvisualization}.
The SENSEI Python \insitu{} component includes three methods that are the locations for user-supplied code: Initialize(), Execute(), and Finalize().

One of the Python-based components -- the bandpass filter -- ingests data from the FFT endpoint, performs filtering by selectively zeroing out certain frequency amplitudes, then produces data for use by the downstream components.
Bandpass filtering is a common way to perform frequency-based filtering where spatial- or temporal-domain data is first converted to the spectral domain, then the unwanted frequencies are eliminated, and then the spectral data is converted back to spatial- or temporal-domain data~\cite{Shenoi:2005}.

The other Python-based endpoint produces a visualization of the filtered data.
Our implementation of this example makes use of Matplotlib's \texttt{imshow} method that is invoked from inside the \texttt{Execute{}} method of the SENSEI Python component.
\section{Results}
\label{sec:results}

%\fix{Wes to make another pass over this section to bring it to completion.}

To demonstrate our implementation, we present an example multi-stage SENSEI \insitu{} workflow that uses heterogeneous components written in C++ and Python.
The workflow makes use of the SENSEI FFT endpoint (\S\ref{sec:design:fft_endpoint}) for performing forward and inverse Fourier transforms on real-valued data and also makes use of two separate SENSEI Python endpoints (\S\ref{sec:design:python_endpoint}) that perform bandpass filtering in the spectral domain and visualization of real-valued results. 

\begin{figure*}
\centering
     \begin{subfigure}[b]{0.49\textwidth}
         \centering
         \includegraphics[width=\textwidth]{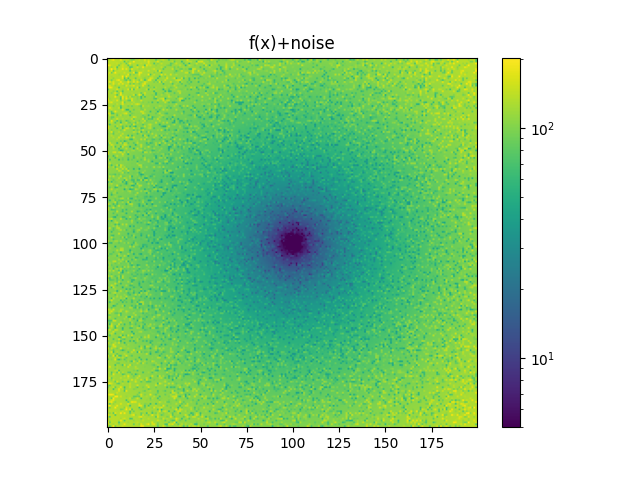}
         \caption{The data generator produces real-valued data in the spatial domain.}
         \label{fig:results:noisy_data}
     \end{subfigure}
     \begin{subfigure}[b]{0.49\textwidth}
         \centering
         \includegraphics[width=\textwidth]{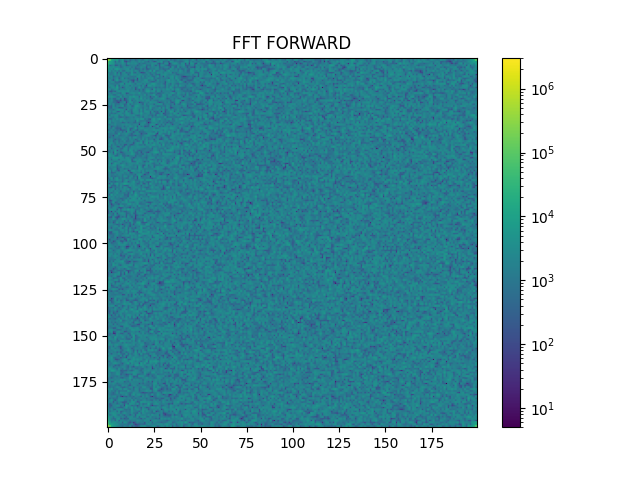}
         \caption{Output from the FFT is spectral-domain data.}
  %       \caption{Using the noisy data as input, output from the FFT is data in the spectral domain indicating amplitudes of different frequencies of the input signal.}
         \label{fig:results:fft_out}
     \end{subfigure}
     \hfill
\centering
     \begin{subfigure}[b]{0.49\textwidth}
         \centering
         \includegraphics[width=\textwidth]{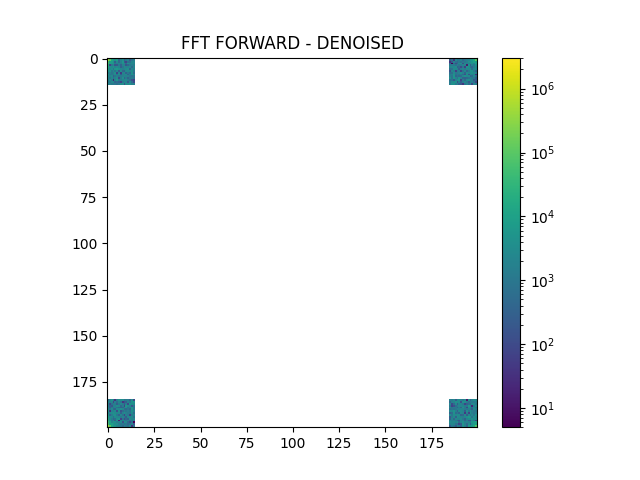}
         \caption{Results of the bandpass filter is also spectral-domain data.}
         \label{fig:results:bandpass_out}
     \end{subfigure}
     \begin{subfigure}[b]{0.49\textwidth}
         \centering
         \includegraphics[width=\textwidth]{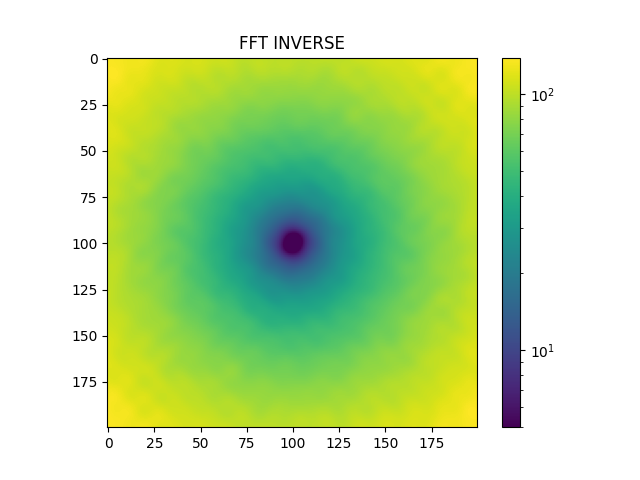}
         \caption{Applying an inverse FFT to spectral data produces real-valued data in the spatial domain.}
         \label{fig:results:fft_inverse_out}
     \end{subfigure}
\caption{These images show visual results at each of the processing stages of the \insitu{} workflow from Fig.~\ref{fig:components_diagram}.}
\label{fig:results}
\end{figure*}

In practice, this type of multi-stage workflow is initiated when the simulation produces data and passes it to SENSEI.
In that case, all the stages of the multi-stage pipeline would execute, and each stage is configured with its own XML file.
In the example we show in Fig.~\ref{fig:components_diagram}, visualization results emerge only after the final stage in the pipeline.
For illustrative purposes, we are visualizing the results of each individual stage in Fig.~\ref{fig:results}.

\subsection{Software Environment}

% Describe the software and versions used in this work.
% What version of SENSEI? What version of FFTW? What was the development platform/platforms? What compilers/OS? 

Along with SENSEI, the FFT backend relies on FFTW library\footnote{\url{https:fftw.org/}} for its core functionality. During this development, the last stable version (3.3.10) of this library is used. This library internally requires a \lstinline{MPI}\footnote{\url{https://www.open-mpi.org/}}  installation.

We have conducted this experiment on \textsc{Perlmutter Cray EX} supercomputer at NERSC at Lawrence Berkeley National Laboratory\footnote{\url{https://docs.nersc.gov/systems/perlmutter/architecture/}}.
Each Perlmutter CPU node houses two AMD EPYC 7763 (Milan) processors with 64 cores per processor clocking at 2.45 GHz. Each node has 512 GB DDR4 memory connected via a memory bandwidth of 204.8 GB/s per CPU. These nodes reach a peak performance of 39.2 GFlops per core.

The software environment consists of a SUSE Linux Enterprise Server 15. We build the program using \textit{GNU compilers version GCC 11.2.0} compiler installed on the servers with the help of \lstinline{'CMake/3.14'} and \lstinline{'make'} tools.

\subsection{Prototype In Situ Workflow}

The demonstration prototype we show here illustrates the use of the new FFT endpoint and two Python-based methods. 
The function implemented by the workflow shown in Fig.~\ref{fig:components_diagram}, is to produce noisy synthetic data, transform the data into the spectral domain using the FFT endpoint, to perform bandpass filtering in the spectral domain to remove high frequencies from the data, inverse transform from spectral back to the real domain, and finally to visualize the results.
%

%A simple prototype illustrated here attempts to use FFT for data denoising.
%For this example, we are producing a small synthetic 2D dataset of $200\times 200$ data points. 
The data source for this demonstration is a custom code that produces spatial data based on a evaluation of a radiating function: \[R = \sqrt{(x - x_c)^2 + (y - y_c)^2}\] where $R$ is the value of point located at $(x,y)$ in the domain centered at $(x_c,y_c)$. 
%
%Figure ~\ref{fig:pure} shows the original dataset.
%
%\begin{figure}[ht]
%    \centering
%    \includegraphics[width=\linewidth]{figures/pure_200x200.png}
%    \caption{Pure data}
%    \label{fig:pure}
%\end{figure}
%
%We will sprinkle some controlled noise over the dataset. 
To this underlying field, we add white noise to about 50\% of the field at random locations.
This output is visualized in Fig.~\ref{fig:results:noisy_data}.

For the sake of this demonstration, the resolution of this data field is quite small: a 2D mesh of dimensions $(200,200)$.
We chose the size of this synthetic dataset so that visualizations of results in Fig.~\ref{fig:results} clearly reveal the characteristics of the data at various stages in processing.
Larger datasets would certainly be more interesting from a visual perspective as well as more useful if the focus of the study is performance benchmarking of the new FFT endpoint.
Here, the primary focus is on functionality, particularly within the context of multi-stage, heterogeneous \insitu{} workflows. 

%For randomly selected half of the total data points, we altered the original value by $\pm 50\%$. This can be clearly seen in Figure ~\ref{fig:noisy}.

%\begin{figure}[ht]
%\centering
%\includegraphics[width=\linewidth]{figures/noisy_200x200.png}
%\caption{Noisy data}
%\label{fig:noisy}
%\end{figure}

The data generator module, shown as \emph{Simulation} in Fig.~\ref{fig:components_diagram}, invokes the FFT Endpoint configured to perform forward transform by passing the noisy dataset to produce data in the spectral 
domain. The output of this pipeline is displayed in Fig.~\ref{fig:results:fft_out}.
%Figure ~\ref{fig:fft}

%\begin{figure}[ht]
%\centering
%\includegraphics[width=\linewidth]{figures/spectral_FFTW_FORWARD_200x200.png}
%\caption{Forward FFT output - spectral domain}
%\label{fig:fft}
%\end{figure}

Next comes the band-pass filtering setup to remove the insignificant frequencies. 
The basic idea is that we will "zero out" regions of the spectral domain to eliminate noise.
In the case of this 2D FFT output, the regions near the four corners represent the lower frequency amplitudes, while regions near the center correspond to higher frequency amplitudes. 
In this particular example, we want to eliminate high frequency noise while preserving the underlying low-frequency signal.
This approach is a perfect use of spectral-based bandpass filtering to eliminate certain types of signal frequencies~\cite{Shenoi:2005}.
%We notice that the corner values hold high values in the spectral domain. 
We retained only $0.75\%$ of the edge values which hold these significant frequencies. Cleaning out the rest of the frequencies gives us Fig.~\ref{fig:results:bandpass_out}.
%Figure ~\ref{fig:denoise}.

%\begin{figure}[ht]
%\centering
%\includegraphics[width=\linewidth]{figures/spectral_FFTW_FORWARD_200x200_denoised.png}
%\caption{Forward FFT output after filtering}
%\label{fig:denoise}
%\end{figure}

We invoke our FFT endpoint again to compute inverse FFT on the filtered frequencies.
The reseult is a reasonably denoised image as shown in Fig.~\ref{fig:results:fft_inverse_out}
%Figure ~\ref{fig:inverse}. 
%\fix{It randomly stopped showing my inverse FFT output: figures folder, spectral FFTW BACKWARD.}

%\begin{figure}[ht]
%\centering
%\includegraphics[width=\linewidth]{figures/spectral_FFTW_BACKWARD_200x200.png}
%\caption{Inverse FFT output - denoised}
%\label{fig:inverse}
%\end{figure}

\section{Previous Work}

Because the FFT is widely used in many different types of scientific applications, there is a great deal of interest in the HPC community in scalable, efficient implementations.
FFTW~\cite{Frigo:FFTW:2005} is widely used and has been tuned to work on several different architectures.
Vendor implementations like Intel's MKL~\cite{MKL:2023} and NVIDIA's cuFFT~\cite{cuFFT:2023} are highly optimized for specific hardware much in the same way that vendor-provided libraries for numerical linear algebra methods are in widespread use and are staples of scientific computing~\cite{Linpack:2023}.  

Despite the importance of FFT in scientific computing, its presence is not widespread in visualization applications and nearly non-existent in \insitu{} infrastructure.
Applications like MATLAB~\cite{MATLAB} provide an FFT implementation where it finds use in applications like signal processing.
VTK contains an FFT class~\cite{VTK_FFT:2023} which has recently become accessible through ParaView~\cite{ParaView_Filters:2023}.
In principle, one could leverage the Catalyst API~\cite{Catalyst:2023} to access the FFT capability inside of ParaView to achieve an \insitu{} implementation of FFT. Such an approach would likely be limited in terms of scalability, as the underlying VTK implementation appears to be serial, as well as difficulty in getting data out of ParaView for use with \insitu{} methods from other sources.

The FFT is considered a staple computational method and many different groups are focusing on providing highly performant implementations.
The absence of such capabilities in the HPC \insitu{} space provides the motivation for our work.

%Our intention is to make accessible a scalable FFT implementation for use in an \insitu{} setting on HPC platforms.

\section{Conclusion and Future Work}

%\fix{Wes to crank on this section more.}

This work presents the foundational design and implementation effort for an FFT method suitable for use in multi-stage, heterogeneous \insitu{} processing pipelines suitable for use on HPC platforms.
One of the main challenges to solve in leveraging an existing high quality FFT implementation for use in an \insitu{} setting, is the data marshaling needed to transform from one data model into that of the FFT implementation and then back again.
When data models are closely aligned, such as when both data models use structure meshes, it is possible to achieve zero-copy performance with minimal overhead~\cite{Ayachit:SENSEI-SC16:2016}. 

Our demonstration pipeline consists of the FFT endpoint, which encapsulates the FFTW scalable implementation, with Python-based methods that perform bandpass filtering and visualization.
Although the absolute size of the mesh produced by the data generator is modest in this example, the primary focus of the work is on  functionality, particularly within the
context of multi-stage, heterogeneous \insitu{} workflows.

While the implementation we describe here is serial, both of SENSEI and FFTW are known to scale to high levels of concurrency.
Future work will consist of building on this initial implementation to perform the data redistribution needed to map from M simulation ranks to N FFTW ranks, as well as running larger problem sizes and studying performance in various scenarios. 

%adding the additional data decomposition needed to redistribute data from M producer ranks to N consumer ranks so that the FFT runs in parallel.

\bibliographystyle{ACM-Reference-Format}
\bibliography{main}

\begin{comment}

%%
%% The acknowledgments section is defined using the "acks" environment
%% (and NOT an unnumbered section). This ensures the proper
%% identification of the section in the article metadata, and the
%% consistent spelling of the heading.
\begin{acks}
To Robert, for the bagels and explaining CMYK and color spaces.
\end{acks}

%%
%% The next two lines define the bibliography style to be used, and
%% the bibliography file.
\bibliographystyle{ACM-Reference-Format}
\bibliography{sample-base}

\end{comment}

%\clearpage

\end{document}